**Evidence for quasi-two-dimensional superconductivity in infinite-layer nickelates**

Wenjie Sun[1,2]†, Yueying Li[1,2]†, Ruxin Liu[2,3]†, Jiangfeng Yang[1,2], Jiayi Li[1,2], Shengjun Yan[1,2], Haoying Sun[1,2], Wei Guo[1,2], Zhengbin Gu[1,2], Yu Deng[1,2], Xuefeng Wang[2,3] and Yuefeng Nie[1,2]*

[1]National Laboratory of Solid State Microstructures, Jiangsu Key Laboratory of Artificial Functional Materials, College of Engineering and Applied Sciences, Nanjing University, Nanjing 210093, P. R. China.

[2]Collaborative Innovation Center of Advanced Microstructures, Nanjing University, Nanjing 210093, P. R. China.

[3]National Laboratory of Solid State Microstructures, Jiangsu Provincial Key Laboratory of Advanced Photonic and Electronic Materials, School of Electronic Science and Engineering, Nanjing University, Nanjing 210093, P. R. China.

†These authors contributed equally to this work

*Corresponding author: ynie@nju.edu.cn

**Abstract**

After being expected as a promising analogue to cuprates for decades, superconductivity was recently discovered in infinite-layer nickelates, providing new opportunities to explore mechanisms of high-temperature superconductivity[1,2]. However, in sharp contrast to the single-band quasi-two-dimensional superconductivity in cuprates[3,4], nickelates exhibit a multi-band electronic structure[5-14] and an unexpected isotropic superconductivity[15,16] as reported recently, which challenges the cuprate-like picture in nickelates. Here, we show the superconductivity in nickelates is actually anisotropic and quasi-two-dimensional in nature, as that in cuprates. By synthesizing high-quality lanthanide nickelate films with enhanced crystallinity and superconductivity ($T_c^{onset}$ = 18.8 K, $T_c^{zero}$ = 16.5 K), strong anisotropic magnetotransport behaviors have been observed. The quasi-two-dimensional nature is further confirmed by the existence of a cusp-like peak of the angle-dependent $T_c$, and a Berezinskii–Kosterlitz–Thouless transition near $T_c$. Our work thus suggests a quasi-two-dimensional superconductivity in infinite-layer nickelates, implying a single-$3d_{x^2-y^2}$-band cuprate-like picture may remain valid in these compounds.

Great amounts of efforts have been made to seek for high-temperature superconductivity (HTSC) beyond cuprates to help understanding the elusive pairing mechanisms[17]. Infinite-layer nickelates have long been considered as a promising candidate owing to the same layered square-planar crystal structure and $3d^9$ electron counts as cuprates[2]. Indeed, superconductivity has been realized in rare-earth nickelates recently[1,18-21], which provides a new platform for understanding the physics of HTSC. As is known, cuprates exhibit anisotropic quasi-two-dimensional (quasi-2D) superconductivity due to their layered structure and quasi-2D electronic structure[3,4], and recent experiments show that HTSC can even exist in monolayer cuprate membranes[22]. As for nickelates, they host many electronic features distinctly from cuprates even though sharing the similar layered structures[5,6]. For example, in contrast to the single Cu-$3d_{x^2-y^2}$ orbital in cuprates[3], infinite-layer nickelates possess multiband Fermi surfaces with quasi-2D hole pockets derived from Ni-$3d_{x^2-y^2}$ orbitals, and one or two three-dimensional (3D) electron pockets mainly from rare-earth-$5d$ orbitals depending on the doping level (Fig. 1a)[5,6,12,13]. Some experimental observations, like the metallic ground states[1,18,19,23] and the missing of long-range magnetic orders[24-26], are proposed to be related to such multi-orbital features[5-14]. Moreover, superconductivity in nickelates displays strikingly unexpected 3D isotropic behavior as reported by recent magnetotransport measurements[15,16], which challenges the cuprate-like single-band picture in nickelates[5,27,28]. As such, whether nickelates are ideally analogous to cuprates is still a question under hot debate and the clarification of above puzzles will greatly contribute to the understanding of superconductivity in infinite-layer nickelates.

Here, we report the synthesis of $La_{0.8}Sr_{0.2}NiO_2$ thin films on $SrTiO_3$ substrates with improved crystalline quality and enhanced superconductivity by reactive molecular beam epitaxy (MBE). Transport measurements show an anisotropic upper critical field $H_{c2}$, a cusp-like peak of angle-dependent $T_c$ under external magnetic field, and a Berezinskii–Kosterlitz–Thouless transition (BKT transition) near $T_c$. All these experimental observations suggest the single-band quasi-2D nature of superconductivity in nickelates, similar to cuprates.

It is well known that the synthesis of high-quality superconducting infinite-layer nickelates are extremely challenging, and all growth and reduction parameters require careful optimizations[1,23,29,30]. Using reactive MBE, we synthesize high-quality lanthanide nickelate films for this work since they have lower lattice mismatch with $SrTiO_3$ substrates (1.25%) compared with neodymium- and praseodymium-based counterparts[23,31]. Reflection high-energy electron diffraction (RHEED) and x-ray diffraction (XRD) characterizations confirm the high crystalline quality of the as-grown $La_{0.8}Sr_{0.2}NiO_3$ films and final reduced $La_{0.8}Sr_{0.2}NiO_2$ infinite-layer films (Extended Data Fig. 1). More importantly, the cross-sectional scanning transmission electron microscopy (STEM) shows no signature of Ruddlesden-Popper stacking faults (Fig. 1b), a common defect observed in infinite-layer nickelates[16,18,19,30]. Meanwhile, sharp interfaces without obvious cation intermixing can be ascertained through the atomic-resolved energy-dispersive x-ray spectroscopy (EDS) maps (Extended Data Fig. 2d). Temperature-dependent resistivity measurements on these samples reveal a greatly enhanced superconductivity compared to other infinite-layer nickelates grown on SrTiO₃ substrates[18,19,23,31], showing a superconducting transition beginning at ~18.8 K and hitting zero-resistance at ~16.5 K (Fig. 1c), providing a clean system to explore inherent properties in these compounds.

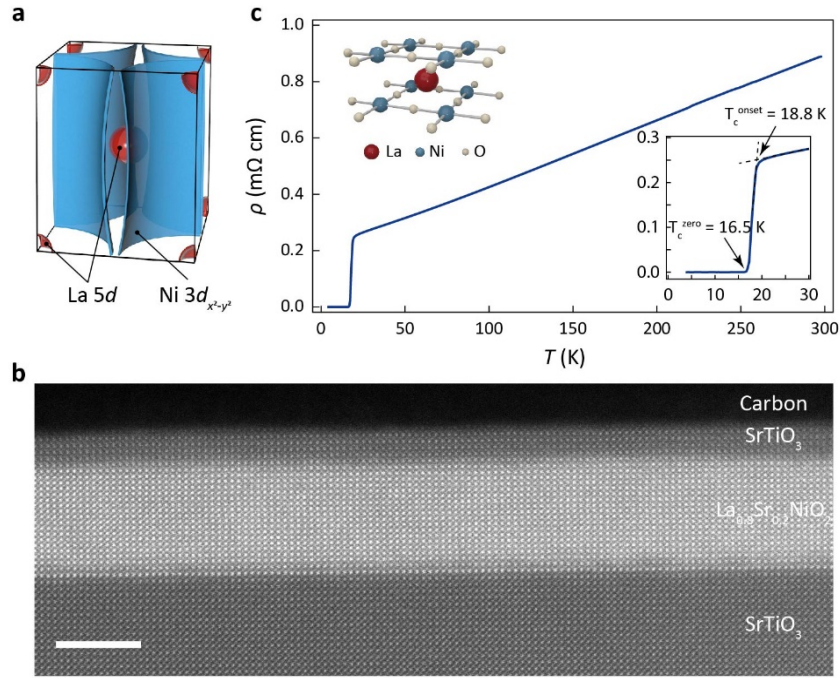

**Figure 1 | Structural and transport characterizations of La₀.₈Sr₀.₂NiO₂ thin film. a,** Schematic drawing of the Fermi surface of infinite-layer nickelates, showing quasi-two-dimensional hole pockets derived from Ni-$3d_{x^2-y^2}$ orbitals, and three-dimensional electron pockets mainly from rare-earth-$5d$ orbitals. **b,** Cross-sectional high-angle annular dark-field scanning transmission electron microscopy (HAADF-STEM) image of a 20-u.c.-thick La₀.₈Sr₀.₂NiO₂ thin film, demonstrating a uniform and defect-free infinite-layer phase. **c,** Temperature-dependent resistivity of La₀.₈Sr₀.₂NiO₂, showing a representative $T_c^{onset}$ and $T_c^{zero}$ of 18.8 K and 16.5 K (bottom-right inset), respectively. The top-left inset shows the crystal structure of infinite-layer lanthanide nickelates,

consisting of NiO$_2$ layers separated by rare-earth cations.

Systematic magnetotransport measurements were performed under external magnetic fields perpendicular ($H \parallel c$) and parallel ($H \perp c$) to the NiO$_2$ plane down to 4 K (Fig. 2a, b). Clearly, the nickelate films display highly anisotropic response to different magnetic field orientations, in sharp contrast to the isotropic behavior reported previously[15,16]. Specifically, in our experiments, superconductivity is robust against the field strength up to 12 T for both field orientations, but shows significantly broader transition width for $H \parallel c$ than $H \perp c$ as the field strength increases. This anisotropic broadening effect of the superconducting transition width is consistent with the field-orientation-dependent vortex motions (see Methods for details).

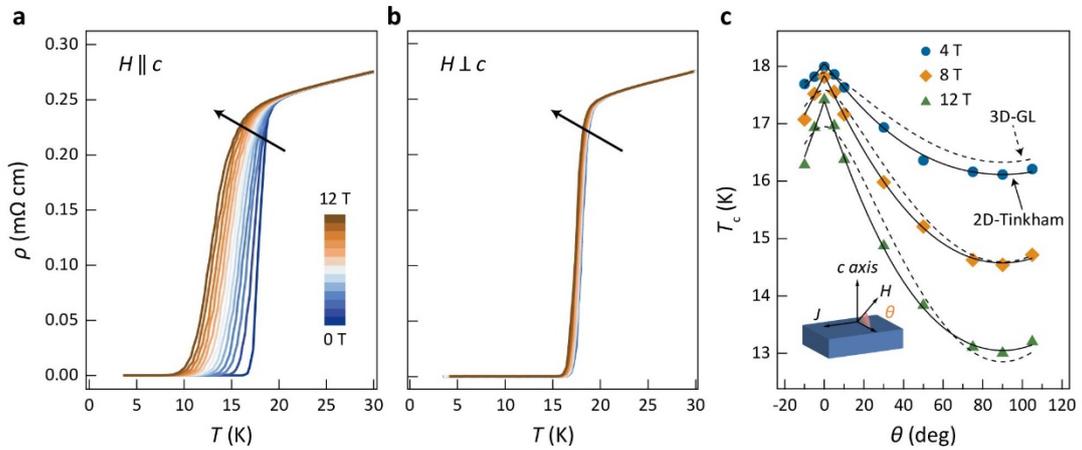

**Figure 2 | Magnetotransport measurements for La$_{0.8}$Sr$_{0.2}$NiO$_2$. a-b,** Resistivity of La$_{0.8}$Sr$_{0.2}$NiO$_2$ thin film under magnetic fields along $c$ axis (**a**) and in the $a$-$b$ plane (**b**) of 0–12 T in increments of 1 T below 30 K. Black arrows denote the increased magnetic field strength. **c,** Angular dependence of the superconducting transition temperature $T_c$ under the magnetic field of 4, 8, and 12 T, fitted with the 2D Tinkham model (solid

lines) and 3D Ginzburg–Landau model (dashed lines). Here and for the rest of the paper, $T_c$ is defined by the criterion of 50% $\rho_n(T)$, where $\rho_n(T)$ is determined through the linear fit to the normal state resistivity between 20 and 30 K. Inset indicates the specific configuration during the angle-dependent measurements, where $\theta$ is the angle between magnetic field and film $a$-$b$ plane.

To quantitatively describe the observed anisotropic behavior, the upper critical field $H_{c2}(T)$ under both field orientations are extracted using the criterion of 50% $\rho_n(T)$ (Fig. 3). Evidently, the $H_{c2}(T)$ displays a linear-dependence for the out-of-plane magnetic fields, while it follows a square-root relation $(T_c - T)^{1/2}$ for the in-plane magnetic fields (inset in Fig. 3a). According to the linearized Ginzburg–Landau (GL) model[32], we obtain the zero-temperature in-plane GL coherence length $\xi_{ab}(0)$ of $27.5 \pm 0.2$ Å and the superconducting thickness $d_{sc}$ of $64.9 \pm 1.9$ Å. Note that the derived $d_{sc}$ is close to the film thickness (~68 Å), indicating the whole film is superconducting, consistent with the high crystalline quality of these films.

As the $(T_c - T)^{1/2}$ dependence of $H_{c2}$ for $H \perp c$ can be originated from the paramagnetic de-paring effects, GL model is inadequate to explain the experimental data since it only takes orbital effects into account[33]. Instead, the single-band Werthamer–Helfand–Hohenberg (WHH) theory[34], which considers both the orbital and paramagnetic de-paring effects, is adopted for the analysis of our $H_{c2}$ data. The overall fitting quality is quite well, yielding the Maki parameter $\alpha_M$ and spin-orbital scattering parameter $\lambda_{so}$ equal to (0, 0) and (56.5, 0.8) for $H \parallel c$ and $H \perp c$, respectively. The essential difference between $\alpha_M$ for two field orientations corresponds well with the anisotropic superconductivity as observed in cuprates[35], that is, orbital-limited for $H \parallel c$ and Pauli-paramagnetic-limited for $H \perp c$. In addition, the relatively large $\lambda_{so}$ signals the spin-orbit effects cannot be ignored here, which may also explain the surpass of the BCS Pauli limit ($H_p = 1.86 \times T_{c,H=0}$) for the case of in-plane magnetic field. Interestingly, the large $\alpha_M$ for $H \perp c$ is in qualitative agreement with previous studies[15,16], which may imply possible existence of the Fulde-Ferrell-Larkin-Ovchinnikov (FFLO) state[36]. Nonetheless, given the system is now in the dirty limit (see Methods for details), further investigations like high-field measurements are needed, which is beyond present work. It should be pointed out that the above fitting analysis are based on the low-field $H_{c2}$ data due to the limited magnetic field strength (up to 12 T), which may lead to quantitatively deviated description of detailed behavior at low temperatures[37]. Nevertheless, the overall anisotropic behavior should be qualitatively reliable as evidenced by the good reproducibility of the fitting results on different samples. (Extended Data Table 1).

Then we performed angle-dependent measurements of $T_c$ under fixed magnetic fields (Fig. 2c), as widely used to confirm dimensionality of superconductivity in 2D materials[38,39] as well as in cuprates[40,41]. As expected, the $T_c(\theta)$ exhibits a strong dependence on the magnetic field orientations, manifesting a strikingly sharp cusp-like feature near $\theta = 0°$ ($H \perp c$). It should be noted that $T_c(\theta)$ extracted using 1% $\rho_n(T)$ and 90% $\rho_n(T)$ criterion both reveal similar behaviors (Extended Data Fig. 4), manifesting that this cusp-like feature is criterion-independent. It is known that the angular dependence of $T_c$ can be described by the Tinkham model for 2D superconductors[41,42]

$$T_c(H, \theta) = T_{c0} - \left|\left(T_{c0} - T_c^{H\|c}(H)\right)\sin\theta\right| - \left(T_{c0} - T_c^{H\perp c}(H)\right)\cos^2\theta \qquad (1)$$

where $T_{c0}$ is the transition temperature at zero-field, $T_c^{H\|c}$ and $T_c^{H\perp c}$ are the transition temperature for out-of-plane and in-plane magnetic fields, respectively. For our data, the cusp-like behavior near $\theta = 0°$ ($H \perp c$) can only be well reproduced by the 2D Tinkham model rather than the 3D Ginzburg-Landau model[42], indicating the quasi-2D nature of the superconductivity in nickelates.

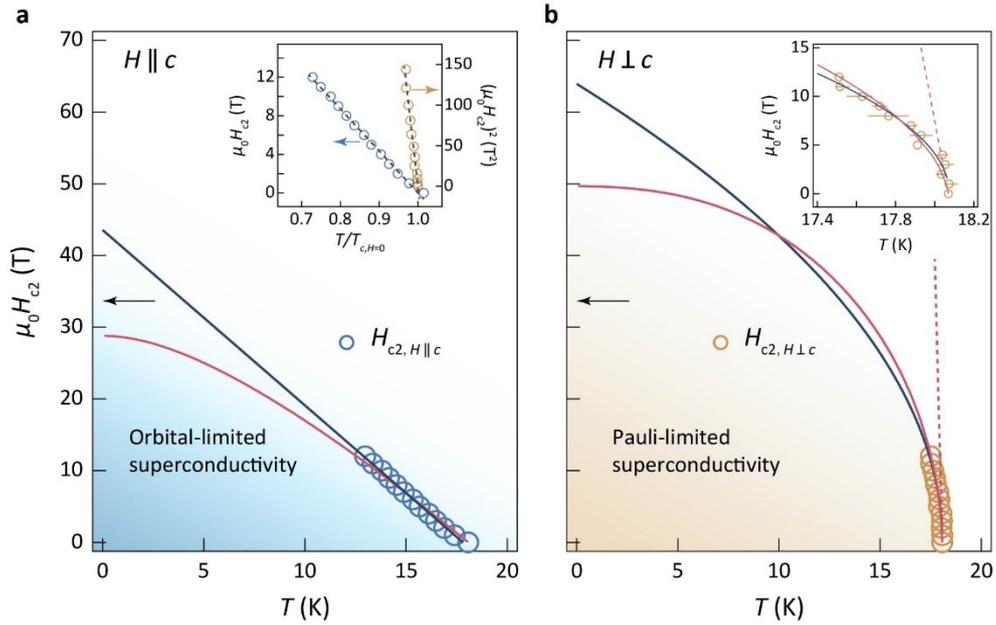

**Figure 3 | $H_{c2}$ versus $T$ phase diagram for La$_{0.8}$Sr$_{0.2}$NiO$_2$. a-b,** Temperature-dependent out-of-plane (**a**) and in-plane (**b**) upper critical field. Here, $H_{c2}$ is defined as the field strength at which resistivity reaches 50% $\rho_n(T)$. The dark-red and the dark-blue solid lines are corresponding fitting to $H_{c2}$ based on the WHH theory and the linearized GL model, respectively. The dark-red dashed line in (**b**) represents the WHH fitting with Maki parameter $\alpha_M$ and spin-orbital scattering parameter $\lambda_{so}$ both equal to zero (pair breaking by orbital effects only). The black arrows denote the BCS Pauli limit of

$H_p = 1.86 \times T_{c,H=0}$ (where $T_{c,H=0}$ is the superconducting transition temperature at $H = 0$ T). The shaded blue and orange area indicate the orbital-limited and Pauli-limited superconductivity, respectively. The inset of (**a**) shows $H_{c2}$ and $(H_{c2})^2$ as a function of the reduced temperature ($T/T_{c,H=0}$) for $H \parallel c$ and $H \perp c$, respectively. The grey dashed lines are the corresponding linear fits. The inset of (**b**) is the enlarged view of the $H_{c2}$ for $H \perp c$ near $T_c$. Error bars represent the temperature uncertainty and are not denoted if they are smaller than the symbol size.

Moreover, we also checked whether the superconducting transition in nickelates can be described by the typical Berezinskii–Kosterlitz–Thouless (BKT) theory for a 2D superconducting system with the transition temperature ($T_{BKT}$) that corresponds to the unbinding of vortex-antivortex pairs[43-45]. According to the BKT model[44], the current-voltage curves of a 2D superconductor show a power-law dependence ($V \propto I^\alpha$) with $\alpha = 3$ at $T_{BKT}$. As illustrated in Fig. 4a and b, the nickelate thin films indeed exhibit a clear BKT behavior and the exponent $\alpha$ reaches 3 at around $T_{BKT} = 17.9$ K, which is almost right at the superconducting transition temperature. In addition, the $R(T) = R_0\exp[-b(T/T_{BKT}-1)^{-1/2}]$ ($R_0$ and $b$ are material-related parameters) relation is expected at the temperature near $T_{BKT}$ according to the Halperin-Nelson's law[46]. As shown in Fig. 4c, the evolution of $R(T)$ is in line with this anticipation yielding $T_{BKT} = 17.8$ K, which agrees well with the value deduced from the power-law analysis. Similar BKT transition behaviors have been widely observed in 2D superconductors, including gallium thin films[47], ion-gated 2D materials[38,39], and cuprates[48].

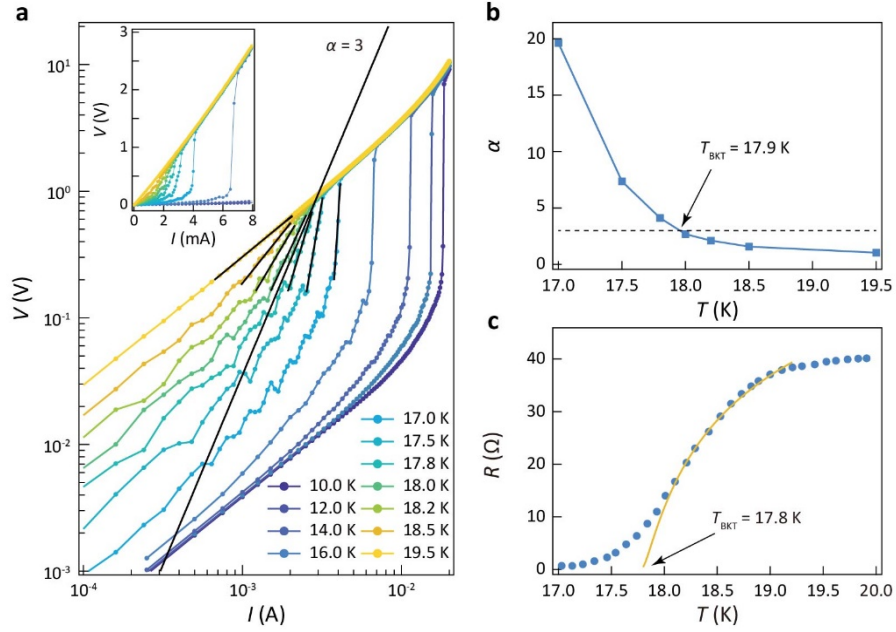

**Figure 4 | Signatures for Berezinskii–Kosterlitz–Thouless transition. a,** Current-voltage characteristics (*I-V* curves) at different temperatures plotted in the double-logarithmic scale. The short black lines are corresponding power-law fits ($V \propto I^{\alpha}$) to the experimental data, while the long black line corresponds to the power-law exponent $\alpha = 3$. The inset shows the same set of *I-V* curves but in the linear scale. **b,** Temperature dependence of $\alpha$, as deduced from the power-law fits shown in (**a**). The interception with the dashed line ($\alpha = 3$) indicates the critical temperature $T_{BKT} = 17.9$ K. **c,** $R(T)$ dependence fitted with a BKT behavior (solid line) as discussed in the main text, yielding a $T_{BKT} = 17.8$ K.

Similar magnetotransport experiments have been performed on many infinite-layer nickelate films, which reproduce the observations (Extended Data Fig. 5 and 6), indicating the superconductivity in nickelates is inherently quasi-2D in nature. And the previous observations of isotropic superconductivity in nickelates may be related to the vertical running defects[15,16]. Naively, trace amounts of conductive filament paths can significantly affect the transport properties. Note that one of these previous studies was done on the $Nd_{0.8}Sr_{0.2}NiO_2$ films grown using our MBE system as well[15]. Compared to $La_{0.8}Sr_{0.2}NiO_2$, the $Nd_{0.8}Sr_{0.2}NiO_2$ films have larger lattice mismatch with $SrTiO_3$ substrates, which leads to lower crystalline quality, as evidenced by the unavoidable extra impurity diffractions in the RHEED patterns during the growth[29].

In summary, we have synthesized high quality infinite-layer lanthanide nickelate thin films using reactive MBE and obtained highest reported superconducting transition temperature on $SrTiO_3$ substrates at ambient condition up to date. The upper critical field $H_{c2}$ is found to be strongly anisotropic, showing the dominating role of the orbital and Pauli paramagnetic effects when magnetic field is applied along the $c$ axis and in the $a$-$b$ plane, respectively. The significant angle-dependent $T_c$ under external magnetic field and the existence of the BKT transition near $T_c$ further confirm the quasi-2D nature of the superconductivity in nickelates. Our findings thus suggest that the quasi-2D hole bands derived from Ni-$3d_{x^2-y^2}$ orbitals may play a more important role than the 3D electron bands in pairing formations in nickelates. As such, the single-band cuprate-like picture may be still valid in describing the superconductivity in infinite-layer nickelates.

**Acknowledgments.** This work is supported by National Key R&D Program of China (Grant No. 2021YFA1400400); National Natural Science Foundation of China (Grant Nos. 1861161004, 61822403, and 11874203) and the Fundamental Research Funds for the Central Universities (No. 0213-14380198).

**Author contributions.** Y.F.N. conceived the idea and supervised the project. W.J.S., Y.Y.L. and S.J.Y. synthesized the perovskite samples with the help of H.Y.S. and W.G.

J.F.Y. and S.J.Y. performed the topotactic reduction experiments. W.J.S. and S.J.Y. characterized the crystalline structure. R.X.L. and X.F.W. performed the transport measurements. Y.Y.L., J.F.Y. and W.J.S. analyzed the transport data. J.Y.L. and Y.D. conducted the STEM measurements. W.J.S., Y.Y.L. and Y.F.N. wrote the manuscript with input from all authors.

**Competing financial interests**

The authors declare no competing financial interests.

**Methods**

**Film growth and sample preparation**

The pristine perovskite $La_{0.8}Sr_{0.2}NiO_3$ films were grown on the $TiO_2$-terminated $SrTiO_3$ (001) substrates by reactive molecular beam epitaxy using a DCA R450 MBE system, with the substrate temperature of 600 °C and an oxidant background pressure of ~1 × $10^{-5}$ Torr using the distilled ozone. During the film growth, reflective high-energy electron diffraction (RHEED) was employed *in situ* to monitor both the growth process and surface quality (Extended Data Fig. 1a). The growth parameters, such as cation stoichiometry, were optimized as reported previously[29]. To obtain the infinite-layer phase, the $La_{0.8}Sr_{0.2}NiO_3$ samples together with ~0.1 g $CaH_2$ powder were sealed in a vacuum chamber with background pressure lower than 1 × $10^{-3}$ Torr, and then heated to ~310 °C for 4 hours, with warming (cooling) rate of 10–15 °C/min.

**Structural characterization**

Specimens for the cross-sectional STEM were fabricated using the focused ion beam (FIB) techniques on a Thermo Scientific Helios G4 X FIB system. The atomic-resolution STEM-HAADF and EDS images were obtained on a double spherical aberration-corrected FEI Titan G2 60-300 system at 300 kV with a field emission gun. To further analyze the lattice constants of the infinite-layer structure, we did the Fast Fourier transform (FFT) (Extended Data Fig. 2b) and the intensity line profile (Extended Data Fig. 2c) of the HAADF-STEM image. In the FFT image, the diffraction spots from the substrate and film align well along $a$ axis while show different periodicities along $c$ axis, manifesting the in-plane coherency and difference on the $c$ lattice constants. The average $c$ lattice constant of the $La_{0.8}Sr_{0.2}NiO_2$ film is 3.46 Å as extracted from adjacent 11 peaks if considering the $c$ lattice constant of $SrTiO_3$ is 3.905 Å, which is consistent with previous reports[23]. X-ray diffraction (XRD) was performed using a Bruker D8 Discover diffractometer, with the Cu-K$_\alpha$ radiation of 1.5418 Å wavelength (Extended Data Fig. 1b, c).

**Transport measurements**

The transport measurements were performed using the standard six-probe configuration at the Cryogen Free Measurement System (CFMS, Cryogenic). The ohmic contacts were made through the ultrasonic aluminum-wire bonding. The magnetotransport measurements were performed by aligning the sample surface parallel or perpendicular to the applied magnetic field via a rotator, while the "parallel" is defined as the position with minimum resistance under applied magnetic fields.

**Upper critical field $H_{c2}(T)$ data fitting**

It is known that the superconductivity can be destroyed when sufficiently high magnetic field (above $H_{c2}$) is applied. And the presence of $H_{c2}$ is relevant to the orbital limit and the Pauli paramagnetic limit[42]. For the purely orbital-limit scenario, superconductivity is suppressed by the magnetic-field-induced Meissner current. On the other hand, spin alignments of Cooper pairs along the external magnetic fields can also lead to pair breaking, which is known as the Pauli paramagnetic limit. We first consider the linearized Ginzburg-Landau (GL) model[32] based on the pure orbital effect where $H_{c2}$ for $H \parallel c$ and $H \perp c$ can be expressed as

$$H_{c2}^{H \parallel c}(T) = \frac{\Phi_0}{2\pi \xi_{ab}^2(0)} \left( 1 - \frac{T}{T_c} \right) \tag{1}$$

$$H_{c2}^{H \perp c}(T) = \frac{\Phi_0 \sqrt{12}}{2\pi \xi_{ab}(0) d_{sc}} \left( 1 - \frac{T}{T_c} \right)^{1/2} \tag{2}$$

where $\Phi_0$ denotes the magnetic flux quantum, $\xi_{ab}(0)$ is the zero-temperature in-plane GL coherence length, and $d_{sc}$ is the superconducting thickness. Based on the above model, we obtain $\xi_{ab}(0) = 27.5 \pm 0.2$ Å and $d_{sc} = 64.9 \pm 1.9$ Å.

In addition, the single-band Werthamer–Helfand–Hohenberg (WHH) theory[34], which takes both of the orbital limit and Pauli paramagnetic limit into account, is adopted to get insight into the superconducting behavior in nickelates. As reported previously, the nickelate films are in the dirty limit[16]. To confirm that in our samples, we first estimate the carrier mean free path $\lambda_{mfp}$ from the transport measurements according to the formula in ref.[49,50] to learn about the cleanliness of the present system

$$\lambda_{mfp} = \frac{\left( 3\pi^2 \right)^{1/3} \frac{\hbar}{e^2} n^{-2/3}}{\rho} \tag{3}$$

where $\hbar$ is the reduced Planck constant, $e$ is the elementary charge, $n$ and $\rho$ are the carrier density and resistivity, respectively. Here, we obtain the values of $n = 1.734 \times 10^{23}$ cm$^{-3}$ and $\rho = 0.25$ m$\Omega$ cm through Hall measurements at 20 K (near $T_c$), yielding the corresponding $\lambda_{\mathrm{mfp}}$ to be 1.6 Å approximately.

The G-L coherence length $\xi_{\mathrm{ab}}(0)$ is determined to be ~27.5 Å from the GL fitting shown above. At the zero-temperature limit, the $\xi_{\mathrm{ab}}(0)$ could be expressed by the in-plane Pippard coherence length $\xi_0$ in the formulas[51]

$$\xi_{ab}(0) = 0.74\xi_0 \quad \text{(in clean limit)}$$

$$\xi_{ab}(0) = 0.85\sqrt{\xi_0 \lambda_{mfp}} \quad \text{(in dirty limit)}$$

Consequently, the $\xi_0$ is deduced to be 36.7 Å and 638.2 Å for the clean and dirty limit, respectively. Obviously, the $\xi_0$ of 36.7 Å is much larger than $\lambda_{\mathrm{mfp}}$ which violates clean limit assumption, indicating our sample should be in the dirty limit with the $\xi_0$ of 638.2 Å. Thus, the WHH fitting formula for dirty limit could be used in our $H_{c2}(T)$ data analysis[34]

$$\ln\frac{1}{t} = \sum_{\nu=-\infty}^{\infty} \left\{ \frac{1}{|2\nu+1|} - \left[ |2\nu+1| + \frac{\overline{h}}{t} + \frac{(\alpha_M \overline{h}/t)^2}{|2\nu+1| + (\overline{h}+\lambda_{so})/t} \right]^{-1} \right\} \tag{4}$$

where $t = T/T_c$ is the reduced temperature, $\overline{h} = 2eH_{c2}(v_F^2\tau/6\pi T_c)$ is the reduced magnetic field, $\alpha_M$ is the Maki parameter, which represents the relative contribution of the orbital-limited and the paramagnetic-limited $H_{c2}$, and $\lambda_{so}$ is the spin-orbit scattering parameter. According to the numerical results of the WHH formula, the fitting parameter $\alpha_M$ should be agreed with the Maki formula[34,52]

$$\alpha_M = 0.53H^{'} \tag{5}$$

and the reduced magnetic field could be expressed as

$$\bar{h} = 4H_{c2} / (\pi^2 H^{'} T_c) \tag{6}$$

where $H^{'} = (-dH_{c2}/dT)_{T=T_c}$ in the unit of T/K.

In the case of $H \parallel c$, the curve of $(\alpha_M, \lambda_{so}) = (0,0)$ could fit the $H_{c2}(T)$ data well, manifesting the predominantly orbital-limited superconductivity, as shown in the Fig. 3a. In contrast, as shown in the Fig. 3b, the purely orbital-limited fit deviated a lot from the experimental data of $H \perp c$, implying the Pauli paramagnetic effect should be considered for the quenching of superconductivity. The best fit to the data of $H \perp c$ is obtained when both the paramagnetic effect and spin-orbit scattering are taken into consideration, with $(\alpha_M, \lambda_{so}) = (56.5, 0.8)$. Given the $H' = \sim107$ T/K for $H \perp c$ in our sample, the parameter $\alpha_M$ is consistent with the value of 56.7 deduced from the Maki formula.

**Thermally activated vortex motion analysis**

As shown in Fig. 2, the superconducting transitions exhibit clear broadening effect along with the increasing magnetic field, which could be attributed to the energy dissipation caused by the thermally activated vortex motion. Similar effects have also been reported in cuprates[53] and iron arsenides[54].

According to the thermally activated dissipation model in terms of flux creep and flux flow[55], the $\ln\rho$ in the region where the resistivity is below 1% of the normal-state resistivity should be in linear with the reciprocal of $T$, as expressed by the Arrhenius relationship

$$\rho(T,H) = \rho_0 \exp(-U_0 / k_B T) \tag{7}$$

where $U_0$ is the activated energy, also represents the pinning potential.

The Arrhenius plots of our $\rho(T)$ data are shown in the Extended Data Fig. 3a and b. The pinning potential $U_0$ could be extracted from the slope of the linear fitting and has a power-law dependence on the magnetic field ( $U_0 \propto H^{-n}$ ) according to the model, as shown in the Extended Data Fig. 3c. The success of the fits to our data indicates that the vortex motion contributes to the transport behavior of the nickelates. There is one or two orders of magnitude difference between the values of $U_0$ for different field orientations, implying the giant anisotropy of vortex motion. In addition, the power-law exponent $n$ is related to the type of pinning force[56]. Extracted from the fitting results, the values of $n$ are 0.13 and 0.65 for $H \parallel c$ and $H \perp c$, respectively. Compared with previous reports in iron-based superconductors[57] and cuprates[55], the obvious difference of $n$ value here manifests the strong anisotropy of the pinning force in the $La_{0.8}Sr_{0.2}NiO_2$ films.

**Angle-dependent superconducting transition temperature ($T_c$)**

The plots of angle-dependent $T_c$ in Fig. 2c were extracted from the resistivity curves versus temperature measured under fixed magnetic fields along different orientations, as shown in Extended Data Fig. 4 a–c. Angular dependence of $T_c$ extracted following different criterions were also shown in Extended Data Fig. 4d–f. $T_{c,\,90\%}$, $T_{c,\,50\%}$ and $T_{c,\,1\%}$ correspond to the temperature at which the resistivity reaches 90%, 50% and 1% $\rho_n(T)$, respectively, where $\rho_n(T)$ is determined through the linear fits to the normal state resistivity between 20 and 30 K. The high-quality fits to all these sets of $T_c(\theta)$ using two-dimensional (2D) Tinkham model demonstrate the criterion-independent quasi-2D superconductivity in nickelates.

**Reproducibility**

To prove the reproducibility of our main conclusion, we did the same magnetotransport measurements on another $La_{0.8}Sr_{0.2}NiO_2$ film (sample #2), the precursor phase of which was grown at different time from the sample shown in Fig. 2 and 3 (sample #1), but with identical thickness (20 u.c.). Its magnetotransport properties are shown in Extended Data Fig. 5, from which a similar anisotropic response for two field orientations can be observed clearly. We also extract the $H_{c2}$ via the 50%$\rho_n(T)$ criterion and apply the GL theory together with WHH theory to fit the $H_{c2}(T)$, as shown in Extended Data Fig. 6. Similar analysis was performed on the results of sample #2 and the relevant parameters were tabulated in Extended Data Table 1.

**Method References**

**Data availability**

The data that support the findings of this study are available from the corresponding authors on reasonable request.

Extended Data for

**Evidence for quasi-two-dimensional superconductivity in infinite-layer nickelates**

Wenjie Sun[1,2]†, Yueying Li[1,2]†, Ruxin Liu[2,3]†, Jiangfeng Yang[1,2], Jiayi Li[1,2], Shengjun Yan[1,2], Haoying Sun[1,2], Wei Guo[1,2], Zhengbin Gu[1,2], Yu Deng[1,2], Xuefeng Wang[2,3] and Yuefeng Nie[1,2]*

[1]National Laboratory of Solid State Microstructures, Jiangsu Key Laboratory of Artificial Functional Materials, College of Engineering and Applied Sciences, Nanjing University, Nanjing 210093, P. R. China.

[2]Collaborative Innovation Center of Advanced Microstructures, Nanjing University, Nanjing 210093, P. R. China.

[3]National Laboratory of Solid State Microstructures, Jiangsu Provincial Key Laboratory of Advanced Photonic and Electronic Materials, School of Electronic Science and Engineering, Nanjing University, Nanjing 210093, P. R. China.

†These authors contributed equally to this work

*Corresponding author: ynie@nju.edu.cn

**Table of contents**

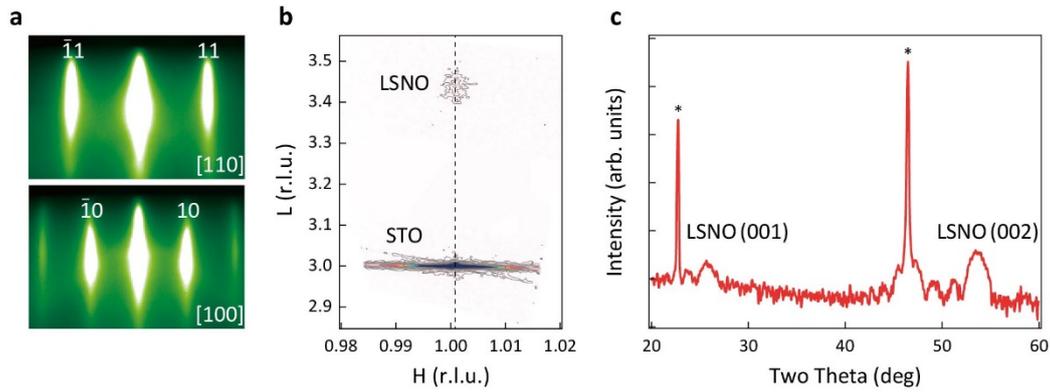

**Extended Data Fig. 1 | Additional structural characterization of nickelate thin films. a**, RHEED patterns along [110] (upper panel) and [100] (lower panel) azimuthal directions taken after the deposition of 20-u.c.-thick perovskite $La_{0.8}Sr_{0.2}NiO_3$ film. **b**, Reciprocal space map of a 20-u.c.-thick $La_{0.8}Sr_{0.2}NiO_2$ thin film around the (103) diffraction peak of the $SrTiO_3$ substrate. The dashed line indicates the nickelate film is fully in-plane constrained to the substrate. **c**, $2\theta$-$\omega$ x-ray diffraction spectrum of the same nickelate film in (**b**). Asterisks denote the diffraction peaks coming from the $SrTiO_3$ substrate.

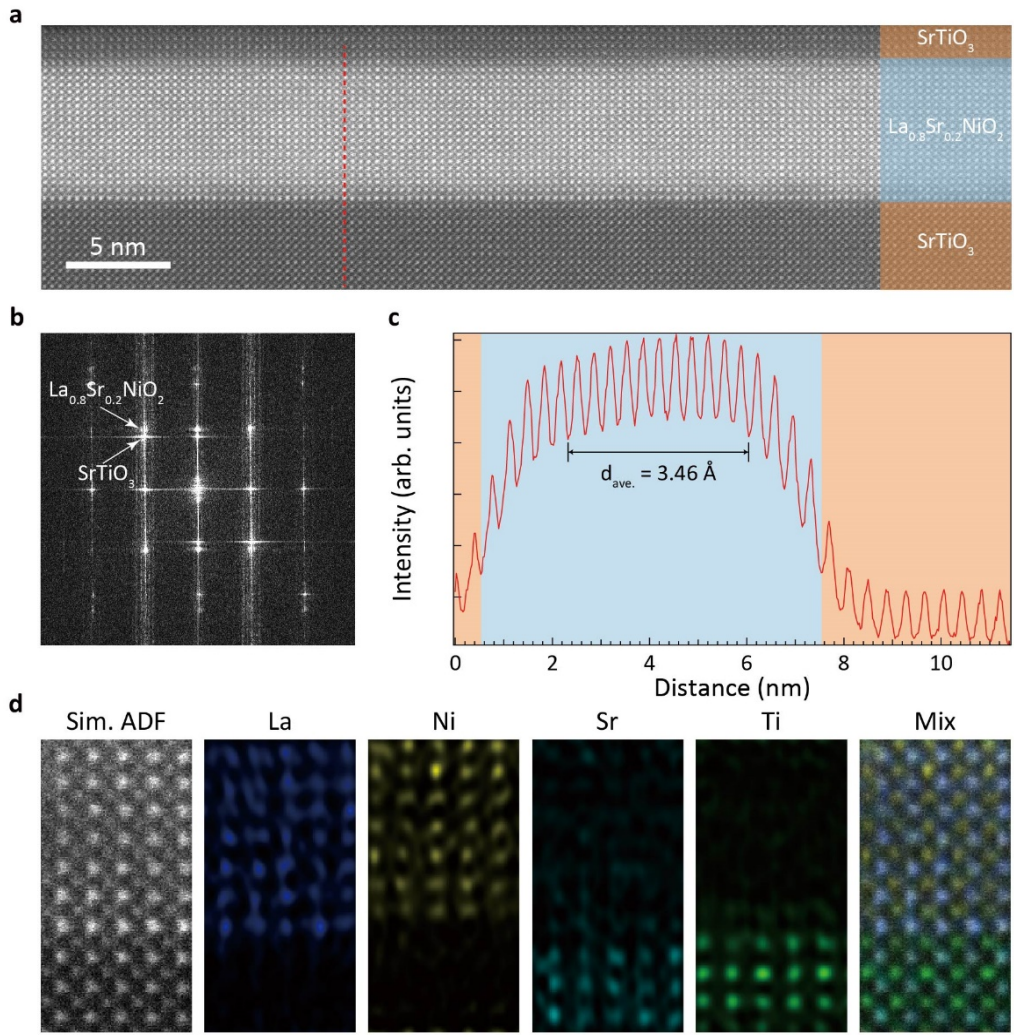

**Extended Data Fig. 2 | Detailed analysis of *c* lattice constants for La$_{0.8}$Sr$_{0.2}$NiO$_2$ thin film. a,** The same HAADF-STEM image of a 20-u.c.-thick La$_{0.8}$Sr$_{0.2}$NiO$_2$ thin film as shown in the main text (Fig. 1b). **b,** Fast Fourier Transform (FFT) image of (**a**), diffraction spots from film and substrate are denoted by white arrows. **c,** Intensity line profile along the red dashed line shown in (**a**). The average *c* lattice constant of the La$_{0.8}$Sr$_{0.2}$NiO$_2$ extracted from adjacent 11 peaks is 3.46 Å if considering the *c* lattice constant of SrTiO$_3$ is 3.905 Å. **d,** Zoom-in view of (**a**) and the corresponding energy-dispersive x-ray spectroscopy (EDS) maps with atomic resolution near the film-substrate interface.

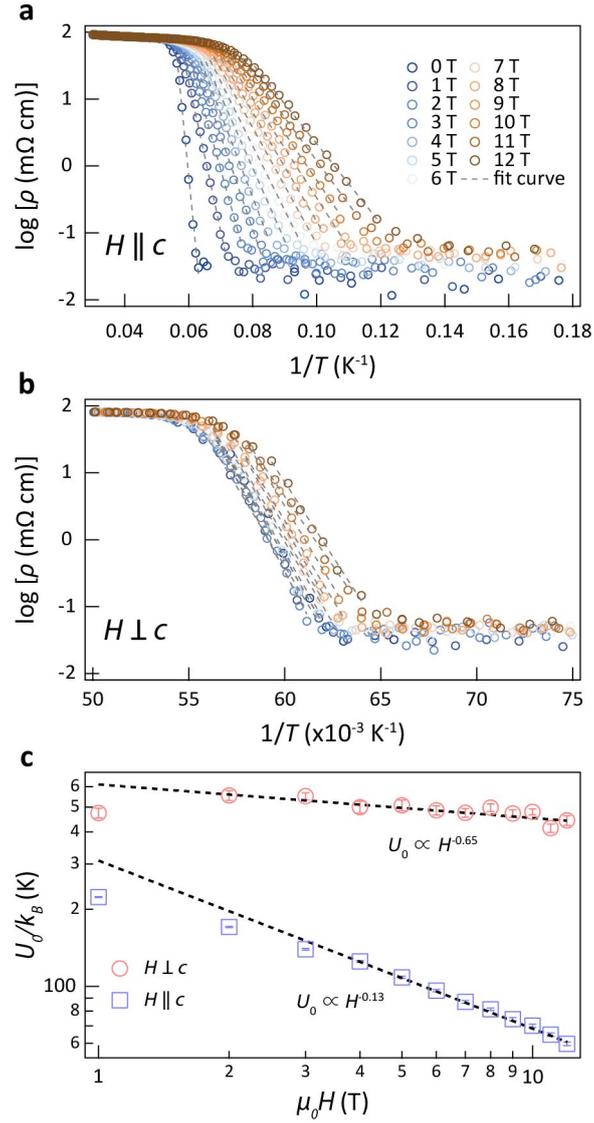

**Extended Data Fig. 3 | Thermally activated behavior of La$_{0.8}$Sr$_{0.2}$NiO$_2$ thin film at different magnetic field orientations. a-b**, Arrhenius plots of temperature-dependent resistivity $\rho(T)$ at magnetic fields along $c$-axis (**a**) and in the $a$-$b$ plane (**b**) of 0-12 T in increments of 1 T below 30 K. **c**, The thermally activated energy $U_0$ extracted by $\rho(T,H)$ = $\rho_0(H)\exp(-U_0/k_BT)$ plotted in double logarithmic scale for both orientations. The dashed lines are the corresponding power-law fits ($U_0 \propto H^n$) to the data above 4 T which yield the exponents $n$ to be 0.65 and 0.13 for magnetic fields along $c$-axis and in the $a$-$b$ plane, respectively.

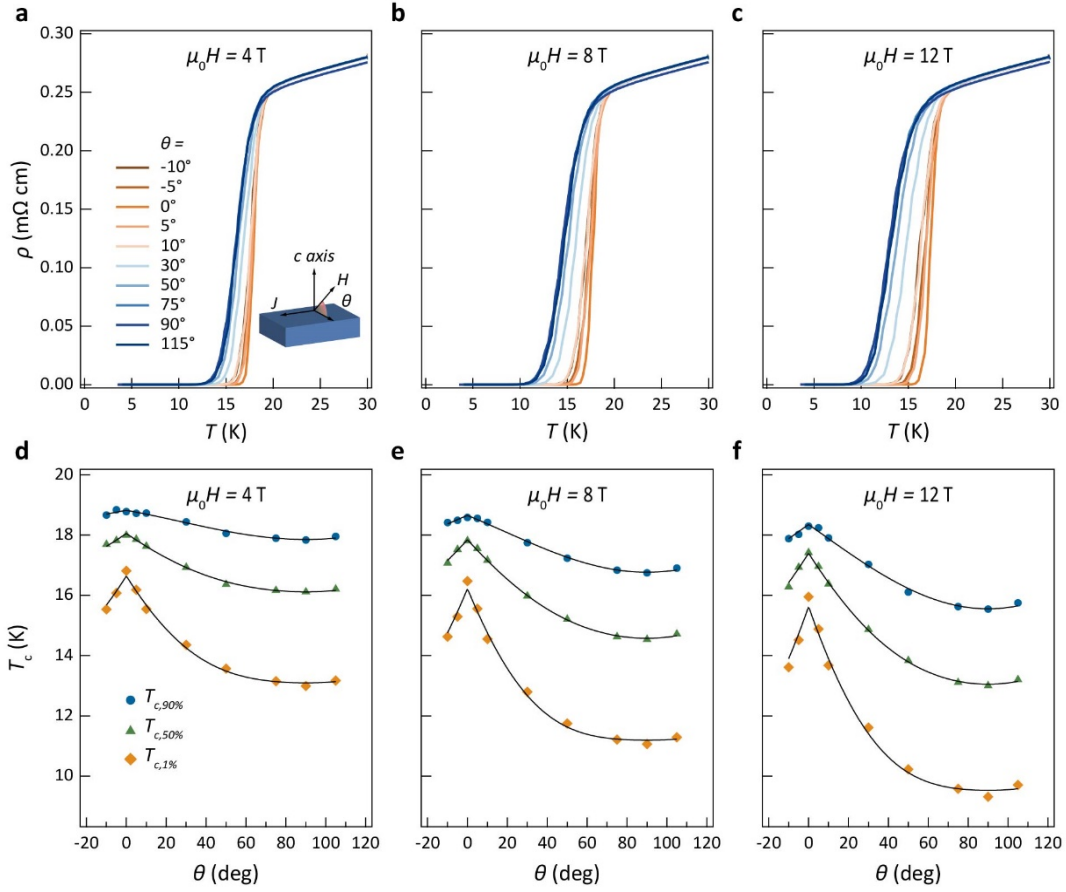

**Extended Data Fig. 4 | Angular dependence of the superconducting transition temperature ($T_c$). a-c,** Temperature-dependent resistivity of La$_{0.8}$Sr$_{0.2}$NiO$_2$ thin film under magnetic fields of 4 T (**a**), 8 T (**b**) and 12 T (**c**) along various orientations. The inset of (**a**) shows a schematic diagram of the configuration. $\theta$ represents the angle between magnetic field and *a-b* plane of the sample. **d-e,** Angle-dependent $T_{c,\,90\%}$, $T_{c,\,50\%}$ and $T_{c,\,1\%}$ of the La$_{0.8}$Sr$_{0.2}$NiO$_2$ thin film under magnetic fields of 4 T (**d**), 8 T (**e**) and 12 T (**f**). The black solid curves are fitting to the experimental data using the two-dimensional (2D) Tinkham model.

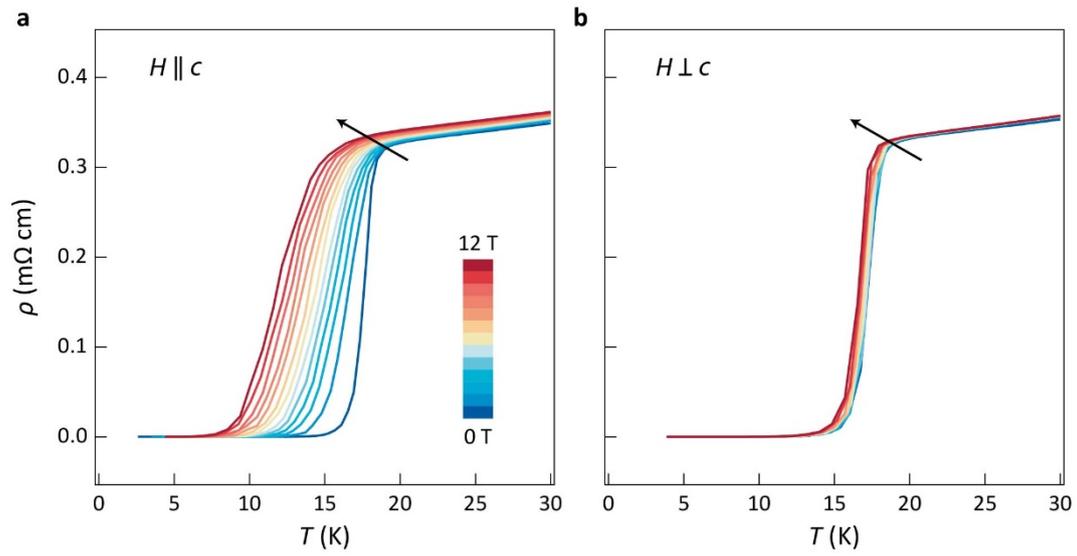

**Extended Data Fig. 5 | Magnetotransport properties for sample #2. a-b,** Resistivity of La$_{0.8}$Sr$_{0.2}$NiO$_2$ thin film under magnetic fields along $c$-axis (**a**) and in the $a$-$b$ plane (**b**) of 0-12 T in increments of 1 T below 30 K. The wave-shaped features of $\rho$-T curves are originated from slightly different acquisition rate of current source meter and temperature controller.

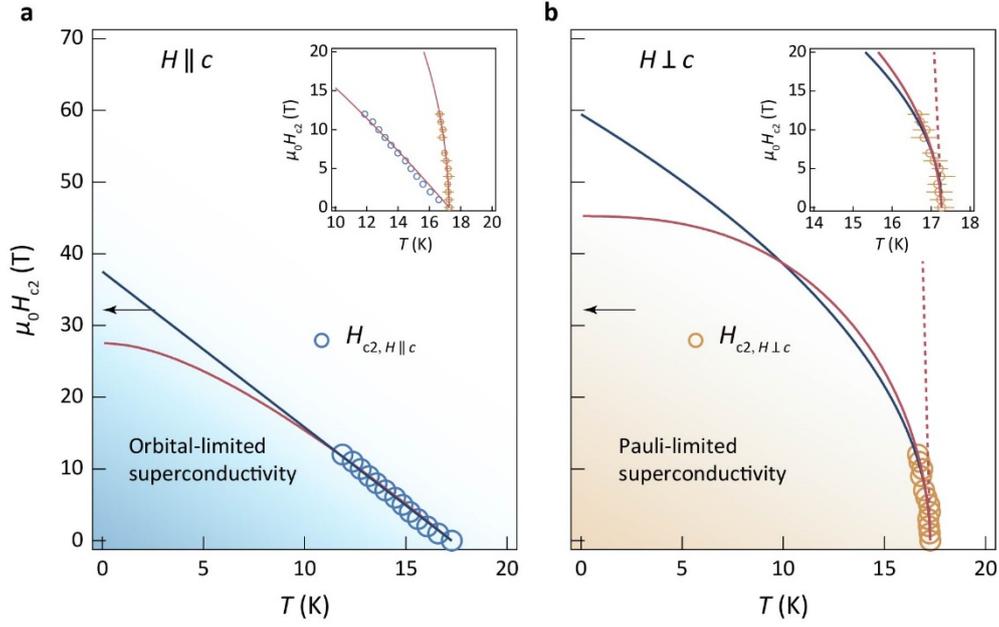

**Extended Data Fig. 6 | The upper critical field $H_{c2}$ for sample #2. a-b,** Temperature-dependent out-of-plane (**a**) and in-plane (**b**) upper critical field for $La_{0.8}Sr_{0.2}NiO_2$. The dark-red and the dark-blue solid lines are calculated $H_{c2}$ values based on WHH theory and GL model, respectively. The dark-red dashed line in (**b**) represents the WHH fitting with Maki parameter $\alpha_M$ and spin-orbital scattering parameter $\lambda_{so}$ both equal to zero (pair breaking by orbital effects only). The black arrows show the BCS Pauli-limit $H_p$ = 1.86 × $T_{c,H=0}$ (where $T_c$ is the superconducting transition temperature at $H$ = 0 T). The inset of (**a**) overlays the $H_{c2}$ data and the corresponding WHH fitting for both filed orientations. The inset of (**b**) is the enlarged view of the $H_{c2}$ data near $T_c$. Error bars of the extracted $H_{c2}$ data represents the temperature uncertainty. Error bars, if not denoted, are smaller than the symbol size.

**Extended Data Table 1 | Summary of the corresponding fitting parameters to $H_{c2}$ of the sample #S1 and #S2.** The subscripts WHH and GL represent the corresponding fitting results deduced from the WHH and GL models, respectively.

| | $T_{c,50\%}$ (K) | $(-dH_{c2}/dT)_{T=Tc}$ (T/K) | $\alpha_M$ | $\lambda_{so}$ | $\mu_0 H_{c2}(0)_{WHH}$ (T) | $\xi_{ab}(0)$ (Å) | $d_{sc}$ (Å) | Film thickness (Å) | $\mu_0 H_{c2}(0)_{GL}$ (T) |
|---|---|---|---|---|---|---|---|---|---|
| #S1 $H \parallel c$ | | $2.3 \pm 0.2$ | 0 | 0 | 28.8 | | | | 43.5 |
| | 18.1 | | | | | $27.5 \pm 0.2$ | $64.9 \pm 1.9$ | 68 | |
| #S1 $H \perp c$ | | $107 \pm 2$ | 56.5 | 0.8 | 49.7 | | | | 63.9 |
| #S2 $H \parallel c$ | | $2.3 \pm 0.2$ | 0 | 0 | 27.5 | | | | 37.5 |
| | 17.3 | | | | | $29.6 \pm 0.2$ | $64.7 \pm 3.5$ | 68 | |
| #S2 $H \perp c$ | | $105 \pm 2$ | 55.5 | 0.7 | 45.3 | | | | 59.5 |